\documentclass{article}
\usepackage{spconf,amsmath,graphicx}
\usepackage[table]{xcolor}
\usepackage{amsfonts}
\usepackage{booktabs}
\usepackage{hhline}
 \pdfoutput=1

\title{Adapting End-to-End Neural Speaker Verification to New Languages and Recording Conditions with Adversarial Training}
\name{Gautam Bhattacharya$^{1,2}$, Jahangir Alam$^{2}$, Patrick Kenny$^{2}$}
\address{$^1$McGill University \\
        $^2$Computer Research Institute of Montreal}

\begin{document}
%
\maketitle
\begin{abstract}

In this article we propose a novel approach for adapting speaker embeddings to new domains based on adversarial training of neural networks. We apply our embeddings to the task of text-independent speaker verification, a challenging, real-world problem in biometric security. We further the development of end-to-end speaker embedding models by combing a novel 1-dimensional, self-attentive residual network, an angular margin loss function and adversarial training strategy. Our model is able to learn extremely compact, 64-dimensional speaker embeddings that deliver competitive performance on a number of popular datasets using simple cosine distance scoring. One the NIST-SRE 2016 task we are able to beat a strong i-vector baseline, while on the Speakers in the Wild task our model was able to outperform both i-vector and x-vector baselines, showing an absolute improvement of 2.19\% over the latter. Additionally, we show that the integration of adversarial training consistently leads to a significant improvement over an unadapted model.
\end{abstract}
\section{Introduction}
\label{sec:intro}

Text-Independent Speaker Verification systems are binary classifiers that given two recordings answer the question: \\
\noindent Are the people speaking in the two recordings the same person?

\vspace{0.5em} The answer is typically delivered in the form of a scalar value or verification score. Verification scores can be formulated as a likelihood ratio, as in the popular i-vector/PLDA approach \cite{p1,p8}. An alternate approach is to use simple distance metrics like mean-squared error or cosine distance. Verification models that can be scored like this typically need to optimize the distance metric itself, i.e. they are optimized end-to-end. While contrastive loss based end-to-end face verification models have shown state-of-the-art performance \cite{p27}, their adoption in the speaker verification community has not been widespread due to the difficulties associated with training such models. 

\vspace{0.5em}\noindent State-of-the-art speaker verification systems follow the same recipe as i-vector systems by using a LDA/PLDA classifier, but replace the i-vector extractor with a Deep Neural Network (DNN) feature extractor \cite{p4}. The DNN embedding model is trained by minimizing the cross-entropy loss over speakers in the training data. While cross-entropy minimization is simpler than optimizing contrastive losses, the nature of the verification problem makes learning a good DNN embedding model challenging. This is evidenced by the Kaldi x-vector recipe, which we use as one of the baseline systems in this work. The recipe involves extensive data preparation, followed by a multi-GPU training strategy that involves a sophisticated model averaging technique combined with a natural gradient variant of SGD \cite{p4}. Replicating the performance of x-vectors with conventional first order optimizers is non-trivial \cite{p28}.

\vspace{0.5em}\noindent In this article we present Domain Adversarial Neural Speaker Embeddings (DANSE) for text-independent speaker verification. We make the following contributions:
\begin{itemize}
    \item We propose a novel architecture for extracting neural speaker embeddings based on a 1-dimensional residual network and a self-attention model. The model can be trained using a simple data sampling strategy and using traditional first order optimizers.
    
    \item We show that the DANSE model can be optimized end-to-end to learn extremely compact (64-dimensional) embeddings that deliver competitive speaker verification performance using simple cosine scoring.
    
    \item Finally, we propose to integrate adversarial training into part of learning a speaker embedding model, in order to learn domain invariant features. To the best of our knowledge, ours is the first to propose the use of adversarial training in a verification setting.

\end{itemize}

\begin{figure*}[htb]
  \includegraphics[width=\textwidth,height=6cm,scale=0.1]{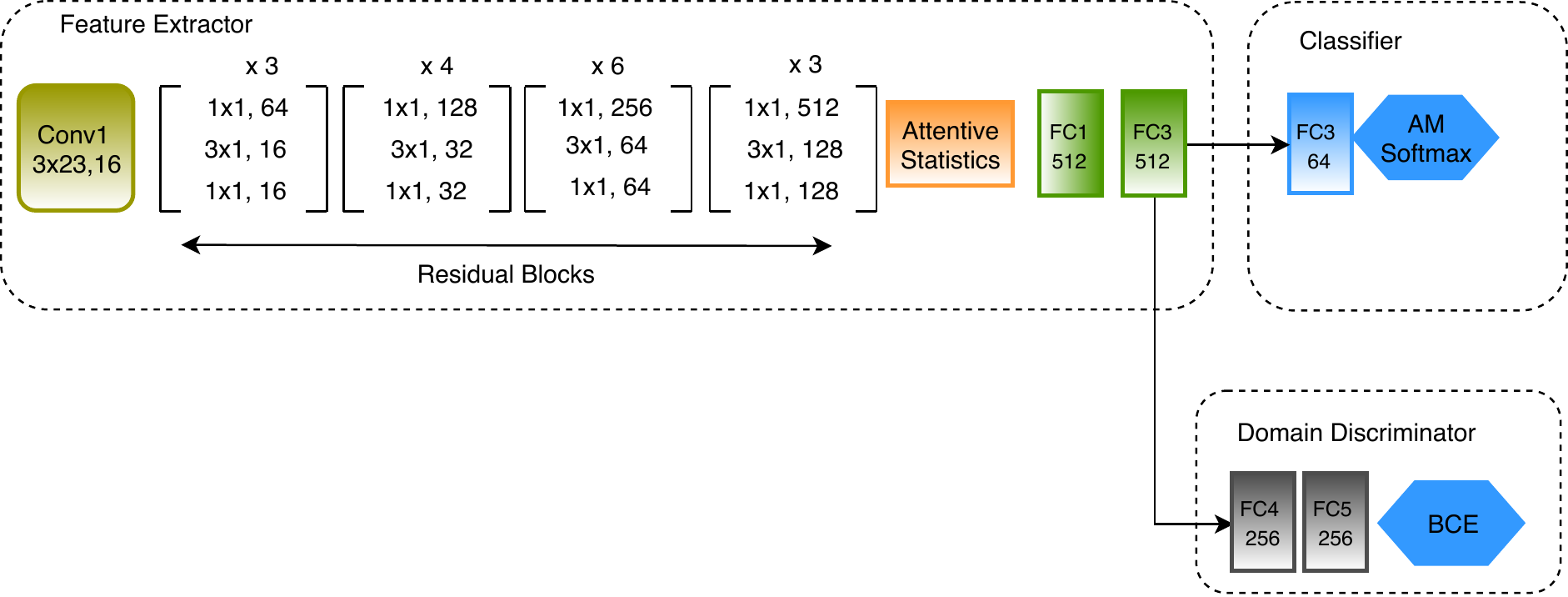}
  \caption{Domain Adversarial Neural Speaker Embedding Model}

  \label{fig:danse}
  
\end{figure*}

\noindent Modern speaker verification datasets like NIST-SRE 2016 and Speakers in the Wild (SITW) are challenging because in-domain or target data is not available for training verification systems \cite{p9,p9a}. This leads to a domain shift between training and test datasets, which in turn degrades performance. Our key insight in this work is that verification performance can be improved significantly by encouraging the speaker embedding model to learn domain invariant features. We achieve this through Domain Adversarial Training (DAT) using the framework of Gradient Reversal \cite{p13}. This allows us to learn domain invariant speaker embeddings using a small amount of unlabelled, target domain data. DAT uses a simple reverse gradient method to learn a symmetric feature space, common to source and target data. This idea has been primarily used to adapt classifiers, but in this work we show that the features learned by DAT are also more speaker discriminative in the target domain. Apart from domain robustness, we find that using an appropriate verification loss function in combination with DAT is equally important for the model to show robust performance using a simple cosine scoring strategy.

\section{Learning Domain Invariant Speaker Embeddings}

 \subsection{Feature Extractor}

The first step for learning discriminative speaker embeddings is to learn a mapping $F(X_{s}) \longrightarrow \textbf{f}$, \ $\textbf{f}\in R^{D}$ from a sequence of speech frames from speaker $s$ to a D-dimensional feature vector $\textbf{f}$.  $F(X)$ can be implemented using a variety of neural network architectures \cite{p4,p5,p6,p7}. In this work we use a deep residual network (ResNet) as our feature extractor \cite{p26}. Motivated by the fact that speech is translation invariant along the time-axis only, we propose to build our model using 1-dimensional convolutional layers. The ResNet architecture allows us to train much deeper networks, and leverage the greater representational capacity afforded by these models. The first convolutional layer utilizes a $3XF$ filter, where $F$ is the dimension of the frequency axis. The residual blocks are followed by an attentive statistics pooling layer (described in next section) and two fully connected layers. In total the feature extractor consists of 52 layers.


\vspace{0.5em}\noindent\textbf{Advantages of ResNet Model}: The main advantage of using a residual architecture is that we are able to learn a very deep speaker representation while maintaining a comparable number of parameters as the baseline x-vector model. Our model has 4.8 million parameters compared to 4.4 million, while our network is over 50 layers deep, while the x-vector network has 7 layers. 

\vspace{0.5em}\noindent Another advantage of the proposed ResNet model is the way incoming audio is processed, which is done at the segment or utterance level. Context information is determined by the size of filter receptive fields, and operations like pooling and striding. In contrast, the baseline x-vector system processes audio at both the frame and segment level, and context is provided through data splicing. As a result, the ResNet model is able to extract speaker embedding much faster than the x-vector system.

\subsection{Self-Attentive Speaker Statistics}

Self-Attention models are an active area of research in the speaker verification community \cite{p7,p24,p25}. Intuitively, such models allow the network to focus on fragments of speech that are more speaker discriminative. The attention layer computes a scalar weight corresponding to each time-step $t$:
\begin{equation}
    e_{t} = \mathbb{v}^{T}f(\textbf{W}h_{t} + \textbf{b}) + k
\end{equation}

\noindent These weights are then normalized, $\alpha_{t}=softmax(e_{t})$, to give them a probabilistic interpretation. We use the attention model proposed in \cite{p24}, which extends attention to the mean as well as standard deviation:
\begin{equation}
        \boldsymbol{\tilde{\mu}} = \sum_{t}^{T} \alpha_{t}\textbf{h}_{t} 
\end{equation}    

\begin{equation}
\boldsymbol{\tilde{\sigma}} = \sum_{t}^{T} \alpha_{t}\textbf{h}_{t}\odot \textbf{h}_{t} - \boldsymbol{\tilde{\mu}} \odot \boldsymbol{\tilde{\mu}}
\end{equation}

 \noindent In this work we apply a self attention model on convolutional feature maps, as indicated in Fig. 1. The last residual block outputs a tensor of size $nBXnFxT$, where $nB$ is the batch size, $nF$ is the number of filters and $T$ is time. The input to the attention layer, $h_{t}$, is a $nF$ dimensional vector. 
 
 By  using a self-attention model, we also equip our network with a more robust framework for processing inputs of arbitrary size than simple global averaging. This allows us simply forward propagate a recording through the network in order to extract speaker embeddings.



\subsection{Classifier}

The classifier block, $C(\textbf{f},\theta_{y})$, is arguably the key component of the model, as it is responsible for learning speaker discriminative features. Recently, angular margin loss functions have been proposed as an alternative to contrastive loss functions for verification tasks \cite{p16}. The Additive Margin Softmax (AM-Softmax) loss function is one such algorithm with an intuitive interpretation. The loss computes similarity between classes using cosine, and forces the similarity of the correct class to be greater than that of incorrect classes by a margin $m$.

\begin{equation}
\begin{split}
    L_{AMS} & = - \frac{1}{n} \sum^{n}_{i=1} \log \frac{e^{s.(cos\theta_{y_{i}} - m)}}{e^{cos\theta_{s.(y_{i}} - m)} + \sum_{j\neq y_{i}} e^{s.(cos\theta_{j})} } \\
    & = - \frac{1}{n}\sum^{n}_{i=1}\log \frac{e^{s.(\textbf{W}^{T}\boldsymbol{f_{i}} - m)}}{e^{s.(\textbf{W}^{T}\boldsymbol{f_{i}} - m)} + \sum_{j\neq y_{i}} e^{s.(\textbf{W}^{T}\boldsymbol{f_{j}})}}
\end{split}
\end{equation}

\noindent Where $\textbf{W}^{T}$ and $\textbf{f}_{i}$ are the normalized weight vector and speaker embedding respectively. The AM-Softmax loss also adds a scale parameter $s$, which helps the model converge faster. We select $m=0.6$ and $s=30$ for all our experiments. 


\subsection{Domain Adversarial Training}
\label{sec:majhead}

So far we have covered the feature extractor $F(X;\theta_{f})$ and classifier $C(\textbf{f},\theta_{y})$ part of our proposed model. In order to encourage our model to learn a symmetric feature space, we augment our network with a domain discriminator $D(\textbf{f},\theta_{d}) \longrightarrow d_{i}$. The discriminator takes features from both the source and target data and outputs the posterior probability that an input feature belongs to the target domain.



\begin{equation*}
\begin{split}
    E(\theta_{f}, \theta_{y}, \theta_{d}) = \mathop{\sum_{\substack{i=1;\\ d_{i}=0}}^{N}} L_{y}(C(F(X_{i};\theta_{f});\theta_{y}),y_{i}) \\
    -  \lambda \mathop{\sum_{i=1...N}}L_{d}(D(F(X_{i};\theta_{f});\theta_{d}),d_{i}) \\
    \hfill (2)
\end{split}
\end{equation*}

Where $\mathbf{L_{y}}$ is the AM-Softmax loss described in section. and $\mathbf{L_{d}}$ is the the binary cross-entropy loss. The objective of domain adversarial training is to learn parameters  $\theta_{f},\theta_{y},\theta_{d}$ that deliver a saddle point of the functional (2):

\begin{equation}
    (\hat{\theta_{f}}, \hat{\theta_{y}}) = \underset{\theta_{f},\theta_{y}}{\operatorname{arg \ min}} \ E(\theta_{f},\theta_{y},\hat{\theta_{d}}) 
\end{equation}

\begin{equation}
    \hat{\theta_{d}} = \underset{\theta_{d}}{\operatorname{arg \ max}} \ E(\hat{\theta_{f}},\hat{\theta_{y}},\theta_{d}) 
\end{equation}

\noindent At the saddle point, the parameters of the domain classifier $\theta_{d}$ minimize the domain classification loss, while the parameters $\theta_{y}$ of the speaker classifier minimize the label prediction loss.  The feature mapping parameters $\theta_{f}$ minimize the label prediction loss - so the features are discriminative, while maximizing the domain classification loss - so the features are domain invariant. The parameter $\lambda$ controls the trade-off between the two objectives \cite{p13}. A saddle point of (5)-(6) can be found using backpropagation:

\begin{equation}
    \theta_{f} \longleftarrow \theta_{f} - \mu_{1} \bigg(\frac{\partial L^{i}_{y}}{\partial \theta_{f}} - \lambda \frac{\partial L^{i}_{d}}{\partial \theta_{d}}\bigg)
\end{equation}

\begin{equation}
    \theta_{y} \longleftarrow \theta_{y} - \mu_{2} \frac{\partial L^{i}_{y}}{\partial \theta_{y}}
\end{equation}
\begin{equation}
    \theta_{d} \longleftarrow \theta_{d} - \mu_{3} \frac{\partial L^{i}_{d}}{\partial \theta_{d}}
\end{equation}

Where $\mu_{1}$,$\mu_{2}$ and $\mu_{3}$ are learning rates. \\
The negative coefficient in eq. (7) induces a reverse gradient that maximizes $L^{i}_{d}$ and makes the features from the source domain similar to those from the target domain. The implementation of the gradient reversal layer is conceptually simple - it acts as the identity transformation during forward propagation, and multiplies the gradient by $-\lambda$ during backpropagation. 

\section{Experimental Setup}
\label{sec:page}


\vspace{0.5em}\noindent \textbf{Training Data} All our systems are trained using data from previous NIST-SRE evaluations (2004-2010) and Switchboard Cellular audio for training the proposed DANSE model as well as the x-vector and i-vector baseline systems. We also augment our data with noise and reverberation, as in \cite{p4}. For speech features extracted 23-dimensional MFCC features from the training set, which mean variance normalization. The baseline i-vector and x-vector systems were trained using the recipies provided with Kaldi. For DANSE model training we filter out speakers with less than 5 recordings.


\vspace{0.5em}\noindent \textbf{Model:} The feature extractor consists of $3XF$ input convolutional layer followed by 4 residual blocks [3,4,6,3], consisting of 48 layers. This is followed by an attentive statistics layer and 2 fully connected layers. The classifier consists of a one hidden layer and the AM-Softmax output layer. The Domain Discriminator consists of 2 hidden layers of 256 units each and the binary cross-entropy (BCE) output layer. We use Exponential Linear Unit (ELU) activations and batch-normalization on all layers of the network.

\vspace{0.5em}\noindent \textbf{Optimization:} We start by pre-training the feature extractor using standard cross-entropy training.  Cross-entropy pre-training is carried out using the RMSprop optimizer with a learning rate (lr) of $0.001$. This learning rate is annealed by a factor of $0.1$ after epochs $4$ \& $8$. We use a simple sampling strategy wherein we define one training epoch as sampling (randomly) each recording in the training set 10 times.

\noindent For training the full DANSE model we found it beneficial to optimize the feature extractor, classifier and domain discriminator with differnet optimizers. The classifier is trained using RMSprop with $lr=0.003$, while the domain discriminator and feature extractor are trained using SGD with $lr=0.001$.  We used performance on held out validation set to determine when to stop training. Gradient Reverasl scaling coefficient $\lambda$ is set to $3.0$ for all experiments.

\vspace{0.5em}{\noindent} \textbf{Data Sampling:} We use an extremely simple approach for sampling data during training. We sample random chunks of audio (3-8 seconds) from each recording in the training set. We sample each recording 10 times to define an epoch. For each mini-batch of source data, we randomly sample (with repetition) a mini-batch from the unlabelled adaptation data for adversarial training.

\vspace{0.5em}\noindent \textbf{Speaker Verification:} At test time we discard the domain discriminator branch of the model, as it is not needed for extracting embeddings. Extraction is done by performing a forward pass on the full recording, and using the 64-dimensional $fc3$ layer as our speaker embeddings. Verification trials are scored using cosine distance. Verification performance is reported in terms of Equal Error Rate (EER).

\section{Results}

\vspace{0.5em}\noindent \textbf{NIST-SRE 2016:} The 2016 edition of the NIST evaluation presented a new set of challenges as compared to previous years. The evaluation data consists of Cantonese and Tagalog speakers. The change in language introduces a shift between the data distributions of the training (source) and evaluation data.


\noindent \textbf{Adaptation Data:} NIST also provides 2272 recordings of unlabelled, in-domain, target data for adapting verification systems. 


\begin{table}[htb]
\begin{center}
\begin{tabular}{*5l}    \toprule
\textbf{\emph{Model}} & \textbf{\emph{Classifier}} &\textbf{\emph{Cantonese}}& \textbf{\emph{Tagalog}}& \textbf{\emph{Pooled}}  \\\midrule
i-vector    & PLDA  & 9.51  & 17.61  & 13.65 \\ \hline
x-vector & COSINE & 36.44 & 41.07 & 38.69\\
\textbf{x-vector} & \textbf{LDA/PLDA} & \textbf{7.52} & \textbf{15.96} & \textbf{11.73}\\ 
x-vector & PLDA & 7.99 & 18.46 & 13.32\\ \hline
AMS & COSINE & 11.44 & 21.22 & 16.28\\

DANSE & COSINE & 8.84  & 17.87 & 13.29\\\bottomrule
 \hline
\end{tabular}
\caption{Performance of different speaker verification systems on NIST-SRE 2016}
\end{center}
\end{table}

\vspace{0.5em}\noindent Table 1. compares the performance of the proposed DANSE model with the baseline i-vector and x-vector systems. The DANSE outperforms the i-vector system, showing a 2.6\% relative improvement in terms of the pooled EER. DANSE performs at the level of x-vectors + PLDA, however we are unable to match the full x-vector + LDA/PLDA recipe. We also see that DANSE outperforms the un-adapted AM-Softmax model by a large margin, indicating the advantage of adversarial training.

\vspace{0.5em}\noindent{\textbf{SPEAKERS IN THE WILD (SITW):}} The SITW database provides a large collection of real-world data with speech from individuals across a wide array of challenging acoustic and environmental conditions. The audio is extracted from open-source video, and while consisting of English speakers (like the training data) there is still a distribution shift due to the difference in the microphones used.  

\noindent \textbf{Adaptation Data:} We use a small random selection of 3000 recordings from the VoxCeleb dataset \cite{p29} as adaptation data. Like SITW, VoxCeleb was also extracted from open-source videos, and hence matches the SITW data more closely than the training data.

\begin{table}[htb]
\begin{center}
\begin{tabular}{*5l}    \toprule
\textbf{\emph{Model}} & \textbf{\emph{i-vector}} &\textbf{\emph{x-vector}}& \textbf{\emph{AMS}}& \textbf{\emph{DANSE}} \\\midrule
\textbf{\emph{EER}}  & 11.47 & 10.51 &  \ \ 9.87  & \textbf{8.32} \\\bottomrule
 \hline
\end{tabular}
\caption{Performance of different speaker verification systems on SITW}
\end{center}
\end{table}

\vspace{-0.5cm}\noindent From Table 2. we see that the DANSE model displays the strongest performance on the SITW dataset, showing a 2.19\% absolute improvement over the x-vector baseline.
 Comparing the performance of our model with and without adversarial adaptation, once again we see a clear advantage for the former, with DANSE outperforming the un-adapted AM-Softmax model by 1.5\%.

\section{Conclusion}

In this work we we presented a novel framework for learning domain-invariant speaker embeddings. By combining a powerful deep feature extractor, an end-to-end loss function and most importantly, domain adversarial training we are able to learn extremely compact speaker embeddings that deliver robust verification performance on challenging evaluation datasets. In future work we will explore other forms of domain adversarial training based on Generative Adversarial Networks \cite{p30}. We will also explore different metrics beyond simple visualization to gain further insight into the feature transformations being induced through adversarial training.

\bibliographystyle{IEEEbib}
\bibliography{mybib}

\end{document}